\documentstyle{l-aa}        
\newcommand{\tief}[1]{_{\rm #1}}

\begin{document}
\thesaurus{03(11.01.2; 11.14.1; 11.19.1; 13.21.1)}
\title{Interpretation of the emission line spectra of Seyfert\,2 galaxies by  
multi-component photoionization models}
\author{Stefanie ~Komossa\inst{1,2} \and Hartmut ~Schulz\inst{1,3}}
\offprints{St.~Komossa (Garching address), skomossa@rosat.mpe-garching.mpg.de}
\institute{Astronomisches Institut der Ruhr-Universit\"at, D-44780
Bochum, Germany \and Max-Planck-Institut f\"ur extraterrestrische Physik,
D-85740 Garching, Germany \and  Radioastronomisches Institut der 
Universit\"at Bonn, Auf dem H\"ugel
71, D-53121 Bonn, Germany}
\date{Received October 7, 1996; accepted December 18, 1996 }
\maketitle
\markboth{St.~Komossa and H.~Schulz: Seyfert\,2 photoionization models}
{St.~Komossa and H.~Schulz: Seyfert-2 photoionization models}
\begin{abstract}
We present multi-component photoionization models allowing for 
{\em local} density inhomogeneities in the NLR
to interpret the emission line spectra of Seyfert\,2 galaxies. 
This approach leads to a successful match of
a large set of line intensities from the UV to the NIR.
In particular, the hitherto elusive NIR features 
[SIII]$\lambda9062+\lambda9531$~ as well as high-ionization lines like
[FeVII]$\lambda6087$~ are                            
consistently fitted. The predictions of
CIII]$\lambda1909$~ and CIV$\lambda1549$~ are considerably improved.

From the detailed analysis of single-component photoionization models
we derive the minimal radial extent of the NLR and the necessary span in density.
Furthermore, we determine constraints on 
suggestions made about the role of matter-bounded
clouds, and on proposed explanations for large [OIII]$\lambda4363/\lambda5007$~
ratios (the so-called `temperature problem'), and assess the usability of some
emission-line ratios as indicators of the ionization parameter. 
We find that a {\em systematic} variation of the cloud column densities
in a population of matter-bounded clouds is inconsistent with the 
trends and correlations exhibited by the emission lines in the diagnostic diagrams.  
Concerning the temperature problem, the only possibility that leads to
an overall consistency with the strengths of all other observed emission lines
is subsolar metal abundances (as compared to e.g. the presence of dust, the 
existence of a high-density component,
or matter-bounded clouds). 

In addition, the consequences of the presence of  
(Galactic-ISM-like) dust internal to the clouds were investigated. 
These models
alleviate the [OIII]-ratio problem 
but did not lead to
overall consistent fits. The most conspicuous fallacy lies in the extreme
underprediction of Fe-lines, which is mainly due to the strong depletion
of the Fe abundance. 

In our final model series, the NLR is composed of a mixture of
metal-depleted ($\sim 0.5~\times$ solar) clouds with 
a radius-independent range in densities  
($10^2$~ to $10^5$~ cm$^{-3}$) distributed
over a range of distances from the nucleus 
(galactocentric radii from at least $\sim$ $10^{20}$~ cm to $10^{21.5}$~ cm, for
$Q\tief{tot} = 10^{54}$~ s$^{-1}$). 
In order to encompass the observed range of each line intensity relative
to H$\beta$, it turns out to be necessary to vary the
spectral energy distribution incident on the clouds,
qualitatively confirming the findings of Ho et al. (1993). We found 
a successful continuum sequence by adding 
an increasing contribution of a hot
black body ($T \approx 200\,000$ K) to a steep powerlaw ($\alpha\tief{uv-x} \approx -2$).
These continua imply that low and high-excitation objects differ in
the strength but not in the basic shape of the EUV bump.
\keywords{Galaxies:
active--galaxies:Seyfert--galaxies:nuclei--Ultraviolet: galaxies}
\end{abstract}
\section{Introduction}
The basic excitation mechanism for the emission lines of Seyfert galaxies is 
generally believed to be photoionization by radiation emerging from the central
power source (e.g. Osterbrock 1989). If this is literally true in the sense that other
heating and ionization mechanisms can be neglected for the formation of most of the
emission-line spectrum, the well-established method of photoionization modeling can be employed
to extract pertinent information about the central continuum source and the
structure of the emission-line region. 
However, the history of AGN photoionizing modeling reveals some   
shortcomings despite a general success in fitting the gross features of the spectra.
 
Photoionization models exploring a {\em wide} range of parameters were first discussed
by Stasi${\acute{\rm n}}$ska (1984a,b). Although presenting mainly single-density
models, she already emphasized the necessity of a multi-component approach.
Ferland \& Osterbrock (1986) and Binette et al.\ (1988) assumed the NLR to consist of a spherical 
system 
of individually optically thin filaments cumulatively becoming optically
thick, with a radial 
dependence of density as $\ n(r) \propto
{r}^{-1}$. In particular, Binette et al.\ noted that black bodies as ionizing
continua are comparatively successful as powerlaws (PL). 

Various
density distributions of the narrow line gas (PL continuum-shape 
and the ionization parameter $U$ kept constant) were studied by Viegas-Aldrovandi
\& Gruenwald (1988). Including relativistic electrons as an additional heating
source they concluded that a combination of these processes
with matter--bounded clouds provides the best fits to observed lines. 
Effects of intra-cloud density stratification were studied by
Binette \& Raga (1990).
Recently, Ho et al. (1993) pointed out that correlations
between certain line ratios could be 
attributed to the influence of a single parameter, the hardness of the
ionizing continuum, parameterized in their models by the powerlaw index $\alpha$.      

Although in many respects successful, these models have a number of deficiencies in
common. 
Whereas models involving powerlaw continua systematically underpredict the stronger 
observed HeII$\lambda$4686 line intensities, 
pure black bodies (or geometrically thick accretion disks in the approximation of
Madau 1988) 
fail to properly account for low ionization 
features like [SII]$\lambda\lambda6716,6731$ (Binette et al.\ 1988,
Acosta-Pulido et al.\ 1990). In addition, all pure 
photoionization models face
the problems of (i) underpredicting the high ionization lines [NeV]$\lambda$3426 and
CIII]$\lambda$1909, CIV$\lambda$1549 (by up to three orders of magnitude) 
whereas [OI]$\lambda6300$ is often reproduced too strong 
and (ii) leading to a too low electron temperature 
(indicated by the ratio [OIII]$\lambda$5007/[OIII]
$\lambda$4363) as compared to observations.  
For those few predictions available for the recently measured 
[SIII]$\lambda\lambda9069,9532$ lines in the near infrared (NIR) spectral region,
no consistency with observations could
be achieved in the sense that the predictions remain too strong
(Osterbrock et al.\ 1992).

The aforementioned difficulties suggest that either the photoionization
approach needs some modifications or 
that other ionizing or heating agents
play a role as e.g. shocks.
Detailed models including shock heating were
presented by Viegas-Aldrovandi and Contini (e.g. 1989 and references therein). 
More recent models for fast shocks that include
the photoionizing effect on the precursor gas of an ionizing continuum, 
which is produced in the shocked region, 
were discussed by Sutherland \&
Dopita (1995). 
Morphological evidence for shocked gas in Seyferts was recently summarized 
by Pogge (1996). 
However, whether these shocks also significantly contribute to the
emission line excitation, is still unclear 
(cf. the critical review by
Morse et al. 1996).  

Here we test the classical assumption of pure
photoionization equilibrium. Some of the pertinent results of the present
study were earlier given in 
Komossa \& Schulz (1994) and Komossa (1993, 1994).
In the meantime photoionization studies were conducted
by Moore \& Cohen (1994,1996)
and by Binette et al. (1996) who both invoke, albeit in different ways, a significant 
contribution of matter-bounded clouds to the emission line spectrum.
They concentrate either on a few selected emission lines plus the
additional information from line profiles
or on a small sample of objects. 

Our approach is to take a large sample
of objects, to use all available emission lines, 
and to relax some classical but possibly doubtful assumptions on the
interstellar-medium-(ISM)-like NLR gas that produces the narrow-line emission. For
instance, rather than assuming a special radial density law to ensure
cloud confinement by a supposed wind, we allow for an
inhomogeneous ISM with a {\em range of densities} at the same galactocentric
radius. {\em A priori}, other parameters are unconstrained as well. 
We have been reinforced to this approach
by the success to find composite models for the extended narrow line region (ENLR) of
NGC\,4151 (Schulz~\&~Komossa 1993). 

The additional freedom leads necessarily to an `explosion' of
the parameter space. In order to narrow down the range of
possibilities we therefore reinvestigate the response of the
line spectrum to the radial extent of the NLR, the gas densities, the metal
abundances, the spectral energy distribution and the cloud column densities.
We also study the consequences of the
inclusion of dust.  
The results are finally used to construct a multi-component model with
the goal to resolve the aforementioned difficulties 
and find the essential parameter(s) that explain(s) {\em the ranges and
correlations} exhibited by the Seyfert\,2 data 
in the various diagnostic diagrams rather than fitting any individual object
that might require some fine-tuning with regard to its own peculiar properties.
\section{Conventions and abbreviations}
Throughout this paper arguments of the log function are the 
physical quantities in cgs units.     
The powerlaw index $\alpha$ is defined
in the sense that the flux distribution $f_{\nu}\propto {\nu}^{\alpha}$.

If not stated otherwise, all line intensities 
are simply denoted by the spectroscopic
designation of the features in question
(in unambiguous cases also the wavelength is omitted),
and the numerical values are given as intensity ratios
relative to H$\beta$.

\section{Data base}
In general, for investigating an object class 
there are two different types of approaches: either one can study 
one or a few well-observed individual objects, with the advantage of a more 
detailed and accurate
treatment but at the risk of misidentifying the truly representative characteristics,
or study a large sample of objects that is usually based on a   
more inhomogeneous data base but allows to utilize the information contained
in collective trends and correlations so that true class properties
might be revealed. 

Here we follow the second approach in order to study group
properties of Seyfert\,2 galaxies.
The necessary measured emission-line ratios from the UV to the NIR
were compiled from 
Ferland \& Osterbrock
(1986), Koski (1978), Shuder \& Osterbrock (1981), Phillips et al.\ (1983), 
Veilleux \&
Osterbrock (1987) and Osterbrock et al.\ (1992). 
A few objects were omitted that had later turned out to have to be differently
classified. The objects chosen for the present study are listed in the
caption of Fig. 1. 
For a homogeneous correction for  
dust reddening we utilized the average galactic extinction curve as listed in Table 7.2 of 
Osterbrock (1989) and an intrinsic value of ${\rm H}\alpha/{\rm H}\beta = 3.1$.
For the remaining 18 out of 37 objects less complete
data were taken from Veilleux \& Osterbrock (1987) and
(for the near-infrared [SIII] and [OII] lines) from Osterbrock et al.\ (1992).
For easy identification in the figures all objects are coded by the numbers given
in Fig. 1. Plots of the most conspicuous
line ratios are presented in Fig.\ 1. For comparison we included the
Seyfert\,1 galaxy NGC\,4151 = No.\ 24. 
%
\begin{figure*}
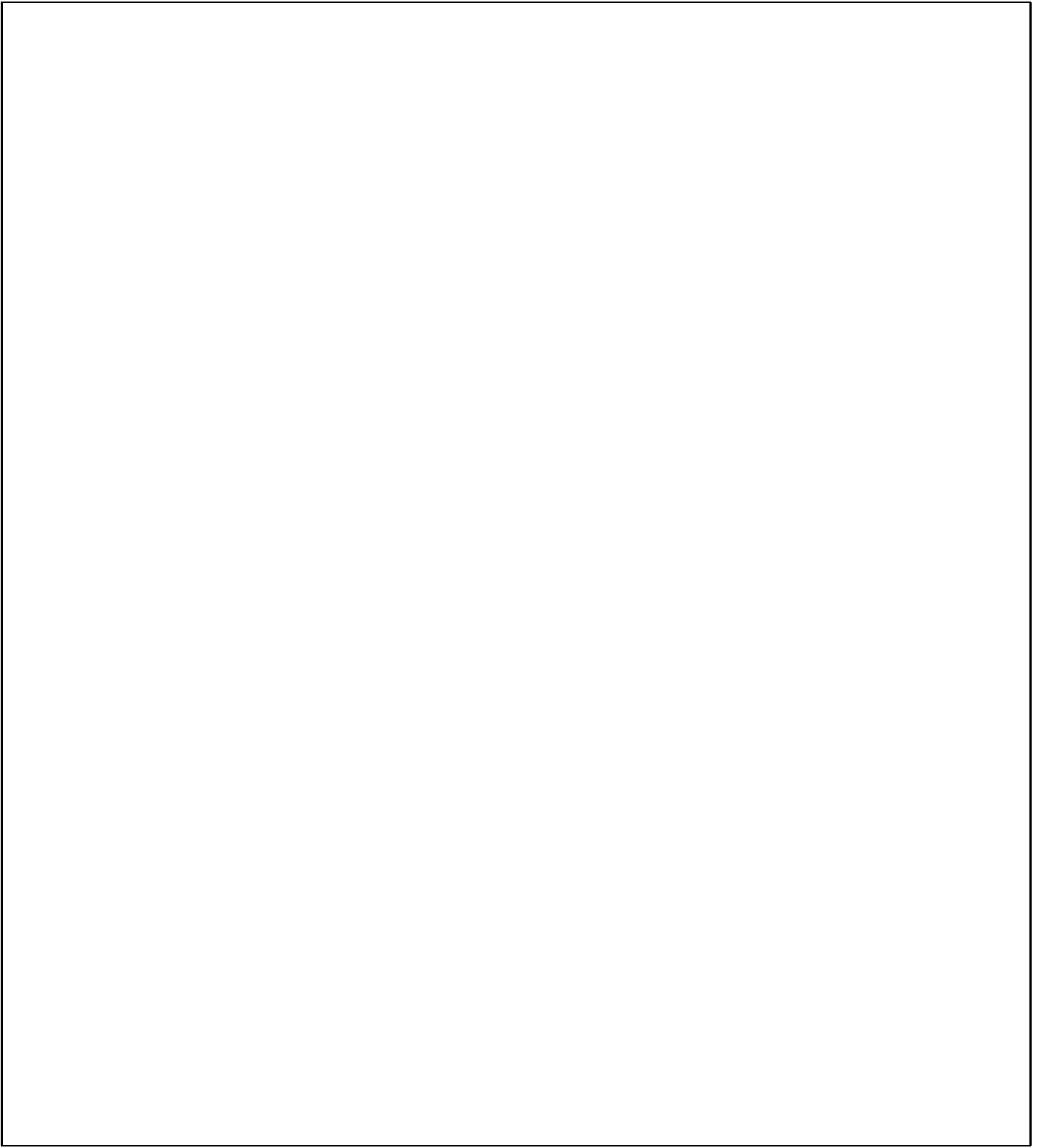

\picplace{20cm}
\caption[ ]{Selection of log (line intensity ratio) diagnostic diagrams.
Numbers correspond to object code as follows:  
1 = Mrk 522, 2 = Mrk 599, 3 = Mrk 1066, 4 = Mrk 273, 5 = NGC 5256, 6 = Mrk 463W,
7 = Mrk 268, 8 =  Mrk 917, 9 = Mrk 198, 10 = III Zw 55, 11 = MrkII 1125+581,
12 = Mrk 463E, 13 = Mrk 1073, 14 = Mrk 270, 15 = MrkII 1133+572, 16 = Mrk 1157,
17 = Mrk 1457, 18 = I Zw 92, 19 = NGC 4388, 20 = NGC 5655, 21 = Mrk 1058, 
22 = Mrk 1388, 23 = Mrk 1, 24 = NGC 4151, 25 = Mrk 34, 26 = Mrk 403, 27 = Mrk 348,
28 = NGC 7674, 29 = Mrk 78, 30 = Mrk 573, 31 = Mrk 3, 32 = NGC 1068, 33 = Mrk 176,
34 = NGC 5643, 35 = IRAS 0147-074, 36 = IRAS 2021+112, 37 = NGC 5506. 
The objects in this sequence are ordered by strength in [OIII]/H$\beta$, with No. 1 
corresponding to the lowest measured value for the sample, No. 37 to the highest. 
}
\end{figure*}

The simplest AGN paradigm assumes that the ionization of the emission-line
clouds is caused by the continuum radiation from a central `point' source.
Assuming in addition, that in the radio to IUE-UV and in the X-ray to
gamma regions the central source is actually observed,
a representative Seyfert continuum was constructed 
by pasting together piecewise powerlaws based on averages taken from 
the literature.
We employed the following slope indices:
in the IR: $\alpha\tief{100\mu-1\mu}=-1$, 
$\alpha\tief{1\mu-0.5\mu}=-2.1$~ (Padovani \& Rafanelli 1988);
in the optical range: $\alpha\tief{opt}=-1.5$~(Koski 1978); 
in the near to far UV:
$\alpha\tief{uv}=-1.4$~(Kinney et al.\ 1991);
in the soft X-ray range:    
$\alpha\tief{sx}=-0.5$~ (Kruper et al.\ 1990);
for harder X-rays:  
$\alpha\tief{hx}=-0.6$~(Awaki et al.\ 1991, Mulchaey et al.\ 1992); 
and a break at 100 keV with $\alpha\tief{}=-3.0$ beyond
(cf. Fig. 2). 
These data fix the continuum
outside the EUV range. However, the region  
between the Lyman limit and 0.5 keV (`EUV') is most important for the ionization.  
In general, this spectral region is
not directly measurable due absorption by neutral gas between observer
and object. For the models, in the EUV range the
flux distribution was varied as described in Sects.\ 4.2.4 and 5.2, and we utilize
the sensitivity of the line reaction on the EUV continuum to constrain the shape of the latter. 
\section{Inferences from single-component
photoionization models}
In order to set the stage for the
multi-component models given in the next section, we first 
give a brief coherent account of the basic effects from the viewpoint
of our one-component models. The calculations were carried out
with version 84.03 of the photoionization code {\em Cloudy} (Ferland 1993).
The program 
calculates the physical conditions in a plane-parallel
slab of gas ionized and heated by the radiation field of a central
object,
by simultaneously solving the equations of statistical and thermal equilibrium. 
The resulting emission line spectrum is self-consistently predicted.   
For details, we refer to Ferland (1993). 
The calculation was stopped either when a pre-specified column
density was reached (matter-bounded clouds) or when the equilibrium
temperature dropped below 4000 K because emission of optical lines
is then negligible (ionization-bounded clouds).
\subsection{Parameter range}
The photoionization models are 
characterized by: \\
\noindent (i) {\em The shape and strength of the ionizing EUV continuum.}
We employ powerlaws
with $\alpha\tief{uv-x}$~from $-1$~to $-2.5$,
black bodies with $T\tief{bb}$~ from 100\,000 to 250\,000 K and 
combinations of both (see Sect. 4.2.4). Examples of the continuum shapes considered are 
displayed in Fig. 2.
%
%
\begin{figure}
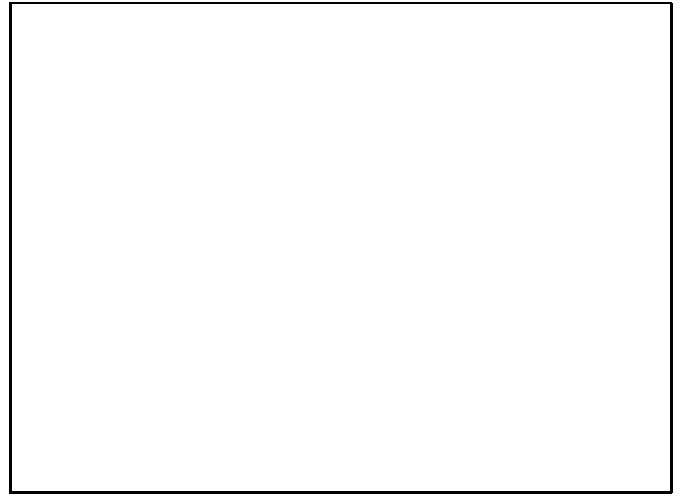

\picplace{6.5cm}
\caption[ ]{Three selected spectral energy distributions used for
the modeling.  Continuous line: black body with $T_{\rm bb} = 160\,000$~ K;
dotted: powerlaw continuum with index 
$\alpha_{\rm uv-x} = -1.5$~ in the EUV; dashed: sum of equal contributions
of an EUV powerlaw
with $\alpha_{\rm uv-x} = -2.0$~ and a black body with 
$T_{\rm bb} = 200\,000$~ K. Slopes for broken
powerlaws outside the EUV region are taken from the literature
(Sect. 3).
 }
\end{figure}

The luminosity of the continuum source is
fixed by the total number rate $Q\tief{tot} = 10^{54}$~s$^{-1}$ of hydrogen 
ionizing photons 
isotropically emitted
by the nucleus.  
Thereby the range of ionization parameters is essentially fixed meaning
that
other values of $Q\tief{tot}$ are also encompassed after scaling the
source--cloud distance range (see Sect. 4.2.1).

\noindent (ii) {\em The geometric and physical properties of the NLR.}
\begin{itemize}
\begin{enumerate}
\item 
As representative distance(s) of the emission-line clouds 
from the central source we employed the four values 
$r_1 = {10}^{20}$ cm, $r_2 = {10}^{20.5}$ cm, $r_3 = {10}^{21}$ cm, 
$r_4 = {10}^{21.5}$ cm; 
\item
We considered as 
hydrogen column densities of the emission-line clouds
$N\tief{H} = {10}^{18...24}~{\rm cm^{-2}}$.
For the adopted hydrogen densities, $N\tief{H} = {10}^{18}$~ cm$^{-2}$ always
corresponds to matter-bounded clouds, ${10}^{24}$~ cm$^{-2}$ 
always to ionization-bounded clouds);
\item  
We systematically examined a range of total  
hydrogen densities $n\tief{H} = {10}^{2...6}~{\rm  cm^{-3}}$~ 
(a number of test calculations were also carried out with
$n\tief{H} = 10^0$~ and $10^1$~ cm$^{-3}$).     
\item
Metal abundances were varied between
0.3 and 3 $\times$~ the solar value (solar abundances
are taken from Grevesse \&
Anders 1989).
\end{enumerate}
\end{itemize}
The combination of (1) and (3) yields the ionization parameter $U$, defined as
\begin{equation}
U=Q\tief{tot}/(4\pi{r}^{2}cn)
\end{equation}
where $n$~ 
= $n_{\rm H}$ is the total hydrogen number density.
Given the range of values of $n\tief{H}$~ and $r$~ chosen,
the models span a wide range of ionization parameters 
between $\log U = -6.58$~ and $+0.42$.
\subsection{Single-component models}
Figure 1 shows some of the diagnostically valuable line ratios. 
The data either form an uncorrelated cloud or correlations 
are apparent (the best examples being [OIII], [NeIII] and
[FeVII] versus HeII (Fig.\ 3)). After including
`high-excitation line' galaxies, Ho et al.\ (1993) found correlations in
the [NII]--[OIII], [SII]--[OIII] and [OI]--[OIII] diagrams. In this section we
explore the possibility whether such trends could be induced by the variation of
a single model parameter (see also discussion in Ho et al.\ 1993)
with special weight given to column-density effects. 
Generally, the effect of a single parameter variation
depends on the choice of the other (fixed) parameters. 
%
\begin{figure*}
\picplace{6.5cm}
\caption[ ]{Observed correlations for selected line-intensity ratios, 
[OIII]$\lambda5007$/H$\beta$, 
[NeIII]$\lambda3869$/H$\beta$, and 
[FeVII]$\lambda6087$/H$\beta$ versus
HeII$\lambda4686$/H$\beta$.
 }
\end{figure*} 

If not stated otherwise, {\em numerical} results given for illustration 
are for the models 
assuming solar elemental abundances
and a `reference continuum' with an EUV powerlaw 
of index $\alpha\tief{uv-x} = -1.5$.    
\subsubsection{Distance of cloud from ionizing source}
$Q\tief{tot}$ and $r_1 \ldots r_4$~ given above correspond to a range
of H ionizing photon fluxes $\Phi = Q/(4\pi r^2)$~ 
from $8 \times 10^{12}$~ phot cm$^{-2}$~ to 
$8 \times 10^9$~ phot cm$^{-2}$~.  The distance range
can be scaled to other values of $Q$~ because $\Phi$~
is decisive as long as geometric
effects
are negligible. 

%
\begin{figure}
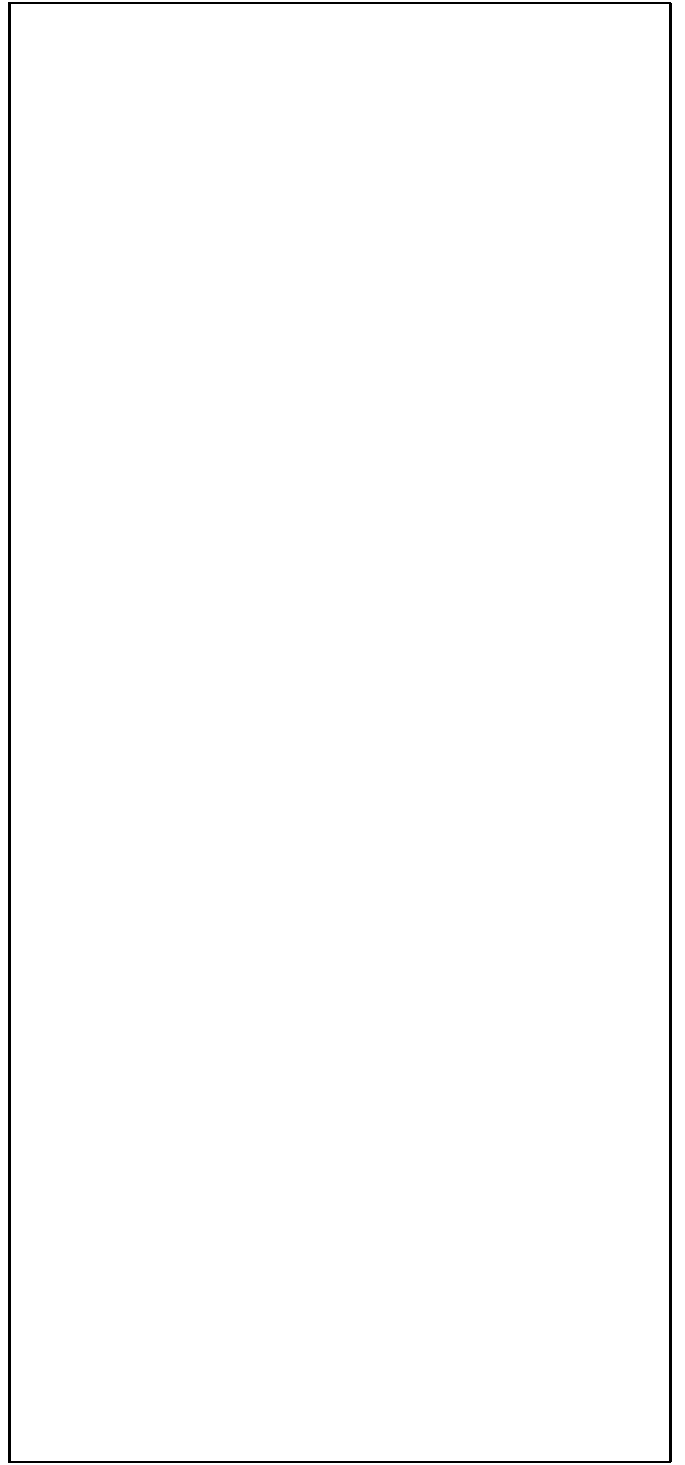

\picplace{19.4cm}
\caption[ ]{Selected emission-line ratios versus hydrogen density. The lines
connect points for constant radial distance $r_i$~ to the central source
(continuous: $r_1={10}^{20}$cm; short-dashed: $r_2={10}^{20.5}$cm; 
long-dashed: $r_3={10}^{21}$cm; dash-dotted: $r_4={10}^{21.5}$cm; cf. Sect. 4.1).
The critical densities for the individual lines are (in cm$^{-3}$) 
$n_{\rm crit}^{\rm [OI]} = 1.8\,10^6$, 
$n_{\rm crit}^{\rm [OII]}$ = 4.5\,10$^3$,
$n_{\rm crit}^{\rm [OIII]}$ = 7.0\,10$^5$, and
$n_{\rm crit}^{\rm [NeV]} = 1.6\,10^7$. 
}
\end{figure}
Fig.\ 4 shows some of the basic results. As might be expected by the
variation of $U$ over the region,
there is a clear trend that low-ionization lines (e.g. from
[OII], [SII], [NI]) are formed at large radii while high-ionization
lines (e.g. CIV, [NeIII], [NeV]) arise closer to the central source.
A comparison with the observations leads to the more stringent
conclusion that {\em all} line ratios cannot be simultaneously reproduced with
clouds at only one distance.   
Moreover, the whole distance range assumed turns out to be necessary. 
E.g., the outermost distance $r_4$ is essential to
model the low-ionization lines which would be too weak for
any mixture of clouds at distances from $r_1$ to $r_3$
regardless of a wide range in density ($\log n\tief{H} = 2-6$),
metal abundance (solar or subsolar) and virtually all 
continua considered.  
\subsubsection{Hydrogen density}
Using standard diagrams (e.g. Osterbrock 1989; his Fig.\ 5.3),
the observed [SII]$\lambda6716/\lambda6731$~ ratios yield 
electron densities of the order $n\tief{e} \approx 10^{2-3}$~cm$^{-3}$, which could,
however, be a misleading average over values from a much wider range of
densities (cf. Mihalszki \& Ferland 1983). Hence, we allow for
hydrogen densities extending from $n\tief{H} = 10^2$~ to $10^6$~ cm$^{-3}$.

Fig.\ 4 
also reveals the pronounced density dependence
(but note that $U$ is not constant). Most lines reach their maximal
strength for $\log n\tief{H} = 3-4$. Others, like [OI] (Fig.\ 4) and, depending
on $r$, MgII$\lambda2798$~ and [SII]$\lambda4072$~  
increase towards higher 
densities and exceed all observed values at $\log n\tief{H} = 6$.

On the other hand, CIV$\lambda1549$~, [NeV]$\lambda3426$~ and further 
high-ionization lines are strong at low densities.
Even for constant $U$ there is a notable density dependence. There is
the expected trend (e.g. Filippenko \& Halpern 1984,
de Robertis \& Osterbrock 1986) of higher emission close to the
critical density (evident from Fig.\ 4 if points
of the same value of $U$ are connected; not drawn). 
However, if one also considers the radius
dependence of the lines it is evident that the implied
$U$ variation over the NLR has a stronger effect than the
closeness of the density to the critical density. This is, e.g.,
demonstrated in Fig.\ 4 by the behavior of [OIII]$\lambda5007$~
and [NeV]$\lambda3426$.

The comparison with the observations reveals that it is neither possible
to fit all observed lines with a single density nor could the observed
trends be explained by a density variation. An appropriate mixture
of components with different densities is required. The amount of the
contribution of a possible high-density component ($\log n\tief{H} = 5-6$) is
limited by the extreme strength of some features like [OI] at these
densities.
\subsubsection{Metal abundances}
We investigated metal abundances between 0.3 and 3 relative to
the solar value. The value of the metal abundance influences the
density of emitting metal ions and the equilibrium temperature. 
The temperature is
important for the efficiency of the collisional excitation. Therefore, changing the 
abundance of a strong coolant affects many lines.
Often the individual variation of the S and N abundances has
no effect on lines from other species. In this case one can
improve the fit of lines from S and N ions by 
tuning the corresponding abundances. In general, the reduction
of metal abundances leads to a lower sensitivity of the emission-line
spectra on the continuum shape.

With suitable parameter choices, most line ratios can be predicted 
to yield acceptable fits for solar metal abundances or even above solar.
The clear exception is 
[OIII]$\lambda5007/\lambda4363$ which is temperature sensitive
below $n\tief{e}\approx 10^5$~cm$^{-3}$~ and therefore signifies that
strong coolants should be suppressed or other heating mechanisms
be introduced. 
\subsubsection{Spectral energy distribution}
For the observable part of the representative continua piecewise
powerlaws were adopted as outlined in Sect. 3. In the
important EUV part we employed a
series of pure powerlaw continua with $\alpha\tief{ux} = -2.5$ to $-1$
and continua with an additional contribution (20\% to 100\% of
the total rate of ionizing photons $Q\tief{tot}$) of
hot black bodies with temperatures in the range 
100\,000--250\,000 K. 
A few
selected continua are shown in Fig.\ 2.

Since the adopted mean continuum has been composed
using average slopes we made test calculations in order to check the
influence of a possible scatter in the observed part (outside the EUV)
of the flux distributions of individual galaxies.
Varying (even to cutting-off) the continuum at wavelengths 
longer than the Lyman limit
has a negligible effect, as expected. Cutting off the total radiation with photon
energies exceeding 0.5 keV leads to a small but insignificant change
of the line ratios. Only a  
strong increase of the X and $\gamma$--contribution 
boosts low-ionization and weakens high-ionization lines, especially for
small radii, small densities and low metal abundances.

In contrast to the observed IR-UV and X-$\gamma$ sections of the
continuum, the shape of the EUV part has a strong influence on the
emission-line spectrum, which, in turn, may be used for its
determination.

Generally, an increase of the slope index $\alpha$ from $-2.5$ to $-1$ in 
powerlaw EUV continua strengthens most line intensities relative
to H$\beta$.   
For solar abundances the strongest
[NII] lines in our sample require $\alpha_{\rm uv-x} = -1$. 
Despite its hardness,
such a flat powerlaw cannot generate most of the strongest 
observed [NeV] and HeII lines. Also, the `canonical' 
powerlaw with $\alpha_{\rm uv-x} = -1.5$~ is not able to  reproduce the total set of
line ratios in the observed ranges.

With black-body EUV continua, 
increasing $T\tief{bb}$ has a similar effect as flattening an EUV
powerlaw. All line ratios relative to H$\beta$ considered here
are strengthened except those of HeI. The strongest HeII, [NeV] and
[FeVII] lines
can be reached. However, the most intense CIII] lines can only be
reproduced with $T\tief{bb} = 250\,000$~ K (at $r_2$~ and for
$\log n\tief{H} = 4$).

If the observed range in HeII is due to a corresponding
range of black-body temperatures then all observed lines
should be correlated with HeII. Some strong lines like [NeIII] and 
[OIII] indeed confirm this expectation, whereas [NI] shows, if any,
a slight anticorrelation
with HeII in our Seyfert\,2 sample. For a larger sample including
more types of emission-line galaxies, Viegas-Aldrovandi (1988) found
an anticorrelation in both [NI] and [OI] with HeII and suggested
as an explanation a systematically increasing contribution of
matter-bounded clouds. We shall pursue this idea in the next section.

Summarizing the results we conclude that it is not possible to explain
the total range of Seyfert\,2 line ratios with a single fixed EUV
continuum shape.
\subsubsection{Column density}

Given the fragmentary structure of the interstellar medium, the
existence of matter-bounded material in the NLR does not appear unlikely.
The knowledge of the presence of such a component is important, e.g.,  
for the determination of the anisotropy of the ionizing continuum
(via the ionization-parameter sensitive [OII]/[OIII] ratio) or for 
estimating temperatures of the emitting gas (via the [OIII]$\lambda$4363/$\lambda$5007 ratio).   

Hydrogen column densities in the range
$N\tief{H} = 10^{18}$~to $10^{24}$~cm$^{-2}$~ were considered to
address the question whether the line-ratio correlations can be
explained by a population of matter-bounded clouds. 
We found
the following difficulties with this scheme:
 
%
\begin{figure}
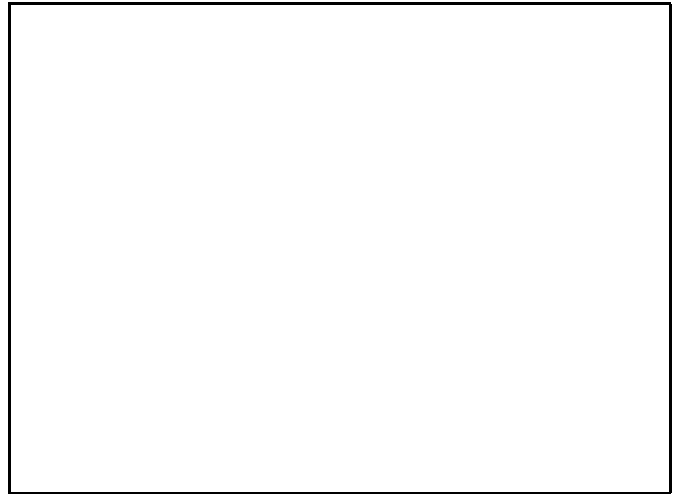

\picplace{6.5cm}
\caption[ ]{Line intensity variations (relative to H$\beta$)
in dependence of column density $N\tief{H}$~ of the emission-line
cloud. The lines are selected for their {\em similar} behavior.
From top to bottom (dashed): [OII]$\lambda$3727, [NII]$\lambda$6583,
[SII]$\lambda$6731, [OI]$\lambda$6300, [NI]$\lambda$5200.  
For comparison, the variation of the total H$\beta$ intensity
(solid line, in arbitrary units) is plotted. 
}
\end{figure}

First, the stronger and, hence, more accurately measured 
[NII], [SII] and [OI] lines are predicted to vary with
column density in essentially the same way as [NI]. However, in our sample 
[NII], [SII] and [OI] are not anticorrelated with HeII
as is claimed for [NI] (cf. Sect.\ 4.2.4). 

Second, there are severe difficulties to produce the [NI]--HeII anticorrelation
by varying $N\tief{H}$, although {\em a priori} such a relation is not unexpected
because HeII increases with lowering $N\tief{H}$~ in contrast to [NI].
However, in order for [NI] to be at least at the lower limit of
its observed strengths, large densities (around 10$^4$ cm$^{-3}$) are
required while a sufficient response of HeII on the column density is only
found for smaller densities. 
For the standard continuum (with $\alpha\tief{uv-x}=-1.5$),  
not even an arbitrary mixture   
of clouds with different
densities, column densities and radial distances could reproduce the
observations. For flatter powerlaw continua ($\alpha\tief{uv-x} \geq -1.5$) or 
hotter black bodies ($T\tief{bb} \ga 150\,000$~K), 
[NI] and HeII become 
stronger. Because it would be extremely contrived and inconsistent
with other line-ratios, we did not
further attempt to look for a mixture of matter-bounded models which matches 
the trend of this single diagram.

%
\begin{figure}
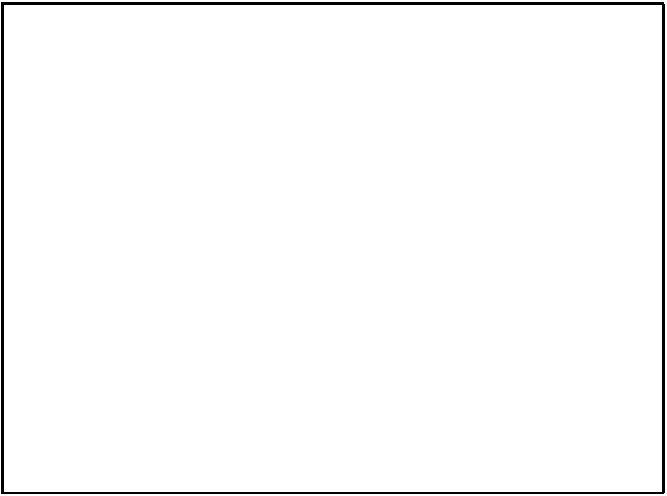

\picplace{6.5cm}
\caption[ ]{Stratification of selected ions in an emission-line cloud,
ion fraction versus cloud thickness $\Delta R$,
which explains the opposite behavior of [NeIII] and the recombination line HeII
in most models.
}  
\end{figure}
Third, any systematic variation of the column, which is sufficiently dominant 
to be seen as a trend in the weak 
[NI] line is expected to have a more profound
effect on stronger lines with lower measurement errors.  
The best positive correlations in our data set are exhibited by  
the strong [OIII] and [NeIII] lines versus HeII (Fig.\ 3). Are these correlations
in agreement with a systematic variation of $N\tief{H}$?
This must be denied because   
there is always an {\em anti}correlation of [NeIII] 
with HeII in the model predictions.   
This is also true
for [OIII] in virtually all cases. 
This kind of behavior can be traced back
to the stratification of the corresponding ions (Fig.\ 6).

In conclusion, the observed correlations between line ratios cannot 
consistently be reproduced by {\em systematically} varying the column density
from object to object.  
Notwithstanding these cautionary remarks on systematic effects,
a random contribution to the line emission by a population of matter-bounded
clouds cannot be excluded and could have introduced some of the 
dispersion in the observations. However, we note that, given the 
typical ionization structure of a NLR cloud, 
a matter-bounded cloud is only expected to significantly contribute
to a given emission line for a very small range in column density 
and strong fine-tuning will be required to explain individual objects 
within such an approach.  
 
\subsubsection{Indicators for the ionization parameter $U$}
As a guide for modeling it would be helpful to have an indicator for $U$~
that is independent on other parameters. We briefly discuss the merits and
deficiencies of two line ratios usable as $U$ indicators.
\paragraph{[OII]$\lambda3727/$[OIII]$\lambda5007$.}
This line ratio has been previously used to 
estimate the value of $U$ (e.g. Penston et al. 1990). 
As shown in Fig.\ 7, independent of
continuum shape and for sufficiently low density ($n\tief{H} < 10^3$~cm$^{-3}$)
this line ratio yields $U$. However, for
higher densities [OII]/[OIII] decreases with $n$ due to the lower
critical density of [OII] ($4.5 \times10^3$~cm$^{-3}$) 
as compared to [OIII] ($7 \times 10^5$~cm$^{-3}$).
If the actual NLR contains a mixture of densities the ratio predicts
too high a value of $U$~ (Schulz \& Komossa 1993). We also note that due to 
different excitation
properties [OII] and [OIII] do not react in the same way on temperature
changes induced by varying the metal or O abundance.

\paragraph{[NeIII]$\lambda3869/$[NeV]$\lambda3426$.}
This ratio is steeply dependent on $U$, but, in addition, 
it is also rather sensitive to the
{\em shape} of the ionizing continuum because [NeV] requires ionizing photons
exceeding 97.1 eV.   
The low sensitivity on density 
makes [NeIII]/[NeV]
a good $U$ indicator also for regions with density inhomogeneities 
provided the continuum shape is sufficiently well known.
%
\begin{figure}
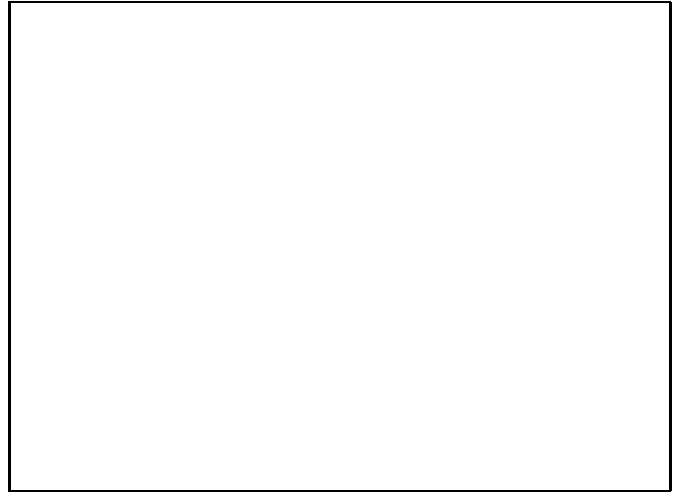

\picplace{6.5cm}
\caption[ ]{Variation of [OII]/[OIII] with the ionization parameter $U$.
Line ratios for constant continuum shape are connected by lines
(continuous: powerlaw $\alpha_{\rm uv-x}=-1.5$; dashed: black body
230\,000 K; dotted: black body 160\,000 K; long-dashed:
black body 100\,000 K). The employed densities
$\log n_{\rm H} = 2, 3, 4, 5, 6$
are marked at the right-hand side of the curves.
}
\vspace{-0.6cm} 
\end{figure}
%

\section{Multi--component models}
As discussed above, a single set of input parameters of one-cloud
photoionization models is neither able to predict the observed strength
of {\em all} observed emission lines nor does a variation of any single 
parameter yield a trend in agreement with the correlations observed.
Consequently, we attempt to build a multi-component model that consists of
clouds with different properties distributed over a certain radial extent.
\subsection{Determination of the weights for radial 
distances and densities}
Since the high-ionization and low-ionization lines are predominantly formed
at small and large radii the full radius range considered here must be 
included (Sects. 4.1 and 4.2.1). The contribution of the innermost radius 
is limited by [NeV] to about 10\% in H$\beta$. After detailed 
inspection of all models 
we finally fixed a 30\% contribution in the H$\beta$ luminosity by gas at each of 
the radii $r_2, r_3$~ and $r_4$~ and a 10\% contribution of the gas at $r_1$.

For this mixture over radius we show the dependence of the most important 
line ratios as a function
of density and continuum shape in Fig.\ 8. It is evident that, for a fixed
continuum, the {\em range} of most observed lines can be covered 
by varying the density. However, it is
not possible to obtain a consistent simultaneous solution of [OIII] and
[OI]. 
%
%
\begin{figure*}
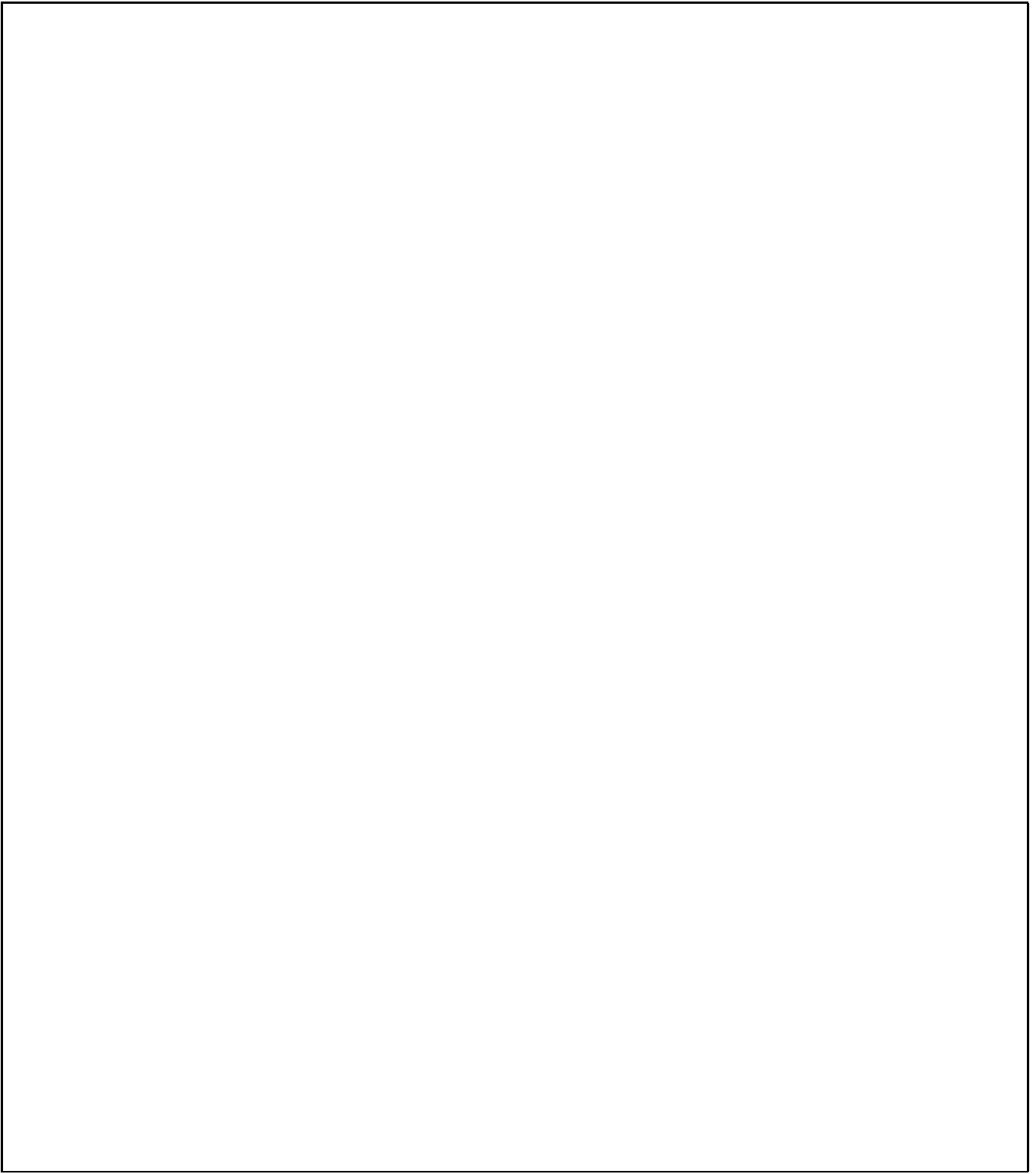

\picplace{20.5cm}
\caption[ ]{These panels show selected diagnostic
diagrams in which models and observations are compared. Open symbols
denote observations. Model emission-line ratios represent a fixed mixture
(see Sect. 5.1) over the four radii $r_1$~ to $r_4$ and are connected by 
a line for each continuum shape. The dotted lines represent black-body
continua with $T\tief{bb} =$ 100\,000 K (labeled `1' in the plots),
160\,000 K and 250\,000 K. The dot-dot-dashed lines refer to powerlaw
continua with $\alpha_{\rm uv-x} = -2.5, -2.0, -1.5$~ (the `reference'
continuum), and $-1.0$. The hydrogen density increases along each  
curve from $10^2$~ to $10^6$~ cm$^{-3}$ (the direction is given by an
arrow on the line of the reference continuum). Filled symbols denote
model results for an additional mixture over density (Sect. 5.1). The
triangles correspond to the powerlaw continua, the circles to the 
black body continua. 
 }  
\end{figure*}
\setcounter{figure}{7}
\begin{figure*}
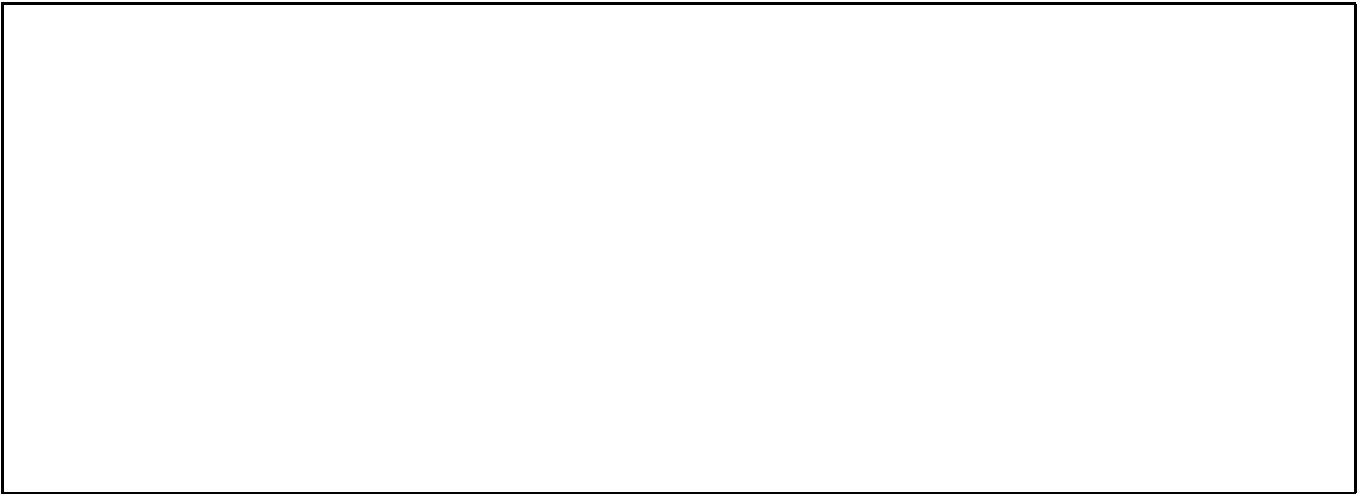

\picplace{6.5cm}
\caption[]{ Continuation. 
}
\end{figure*}

Many lines are satisfactorily represented for $n\tief{H} = 10^{3-5}$~
cm$^{-3}$. However, a contribution of lower densities is necessary because
the [SII] ratio indicates a lower average 
and lines from highly ionized species like CIV and [NeV] call for
low-density gas at smaller radii. 

A radially falling density law, e.g. $n \propto r^{-2}$ so that 
$U = {\rm const}$, can be ruled out for the radius range
considered in this paper e.g. by the strength of [NeV], which requires
a higher than average $U$ at the smaller radii.

From various density distribution functions studied, the
one finally chosen equally weighs the lower hydrogen densities 
($\log n_{\rm H} = 2,3,4$ with weight factors of 0.3) 
to correctly predict at the
same time the high-ionization lines like [NeV] (which demands a high
ionization parameter) as well as the low-ionization lines like [NII],
[SII] (which originate further outwards preferentially at medium densities).
The high-density end is limited by the strength of (then significantly
boosted) [OI] leading to a weight of 0.1 for $\log n_{\rm H} = 5$.  
\subsection{Trends by varying the continuum shape}
Having carried out an additional mixing over densities we are now looking
for a parameter to sweep over the data clouds in the diagnostic diagrams.
It turns out that by appropriate variation of the
continuum shape the observed range of each line ratio is encompassed. 
The following results lead to constraints of the continuum:
models with powerlaw continua cannot reach the strong HeII lines within
the sample, which can, however, be
matched with black body continua. Conversely, black bodies
of the assumed temperature range 
lead to difficulties with the lower part of the [OIII] observations.

These findings suggest to combine a steep powerlaw with an increasingly
dominating contribution of a hot black body. Below we adopt such an
EUV continuum consisting of a powerlaw with $\alpha\tief{uv-x} = -2$~ and
a black body with $T\tief{bb} = 200\,000$~K that contributes
by 0, 20, 50, 80 and 100\% to 
$Q\tief{tot} = Q\tief{pl} + Q\tief{bb} = {\rm const}$.
%
%
\begin{figure*}
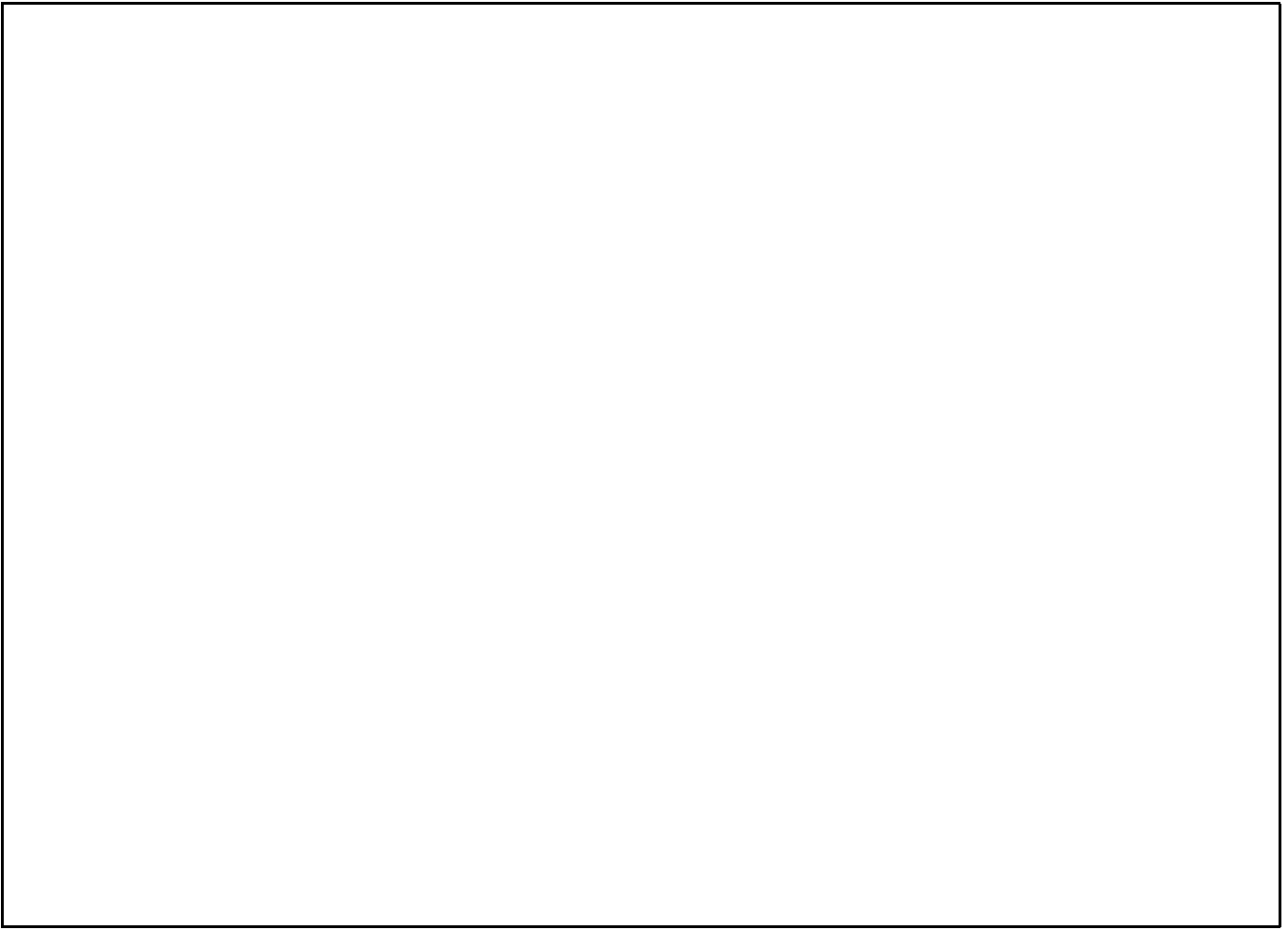

\picplace{13cm}
\caption[ ]{{\bf a}) Dependence of the [OIII]$\lambda 4363/\lambda5007$~ ratio versus
[OIII]$\lambda5007/$[OII]$\lambda\lambda3726,29$ on cloud distance $r_i$~ and
hydrogen density $n\tief{H}$. Continuous, dotted, dashed and dot-dashed
for $r_1={10}^{20}$cm, $r_2={10}^{20.5}$cm, $r_3={10}^{21}$cm, $r_4={10}^{21.5}$cm, 
respectively (Sect. 4.1). Along
each line, $\log n\tief{H}$~ increases from 2 to 6 
(direction indicated by the
arrow). Solar abundances are adopted;
{\bf b}) as in Fig. 9a, except that $\log n\tief{H}$~= 4 is kept
constant and the column density is varied along the lines from
$\log N\tief{H} = 18 \mbox{  to  } 24$;
{\bf c}) as in Fig. 9a, but now after including the effects of dust
and correspondingly depleted elemental abundances. The small dots correspond
to the original (not reddening corrected) observed line ratios;
{\bf d}) as in Fig. 9a, except that the upper series of lines corresponds
to metal abundances of 0.3~$\times$~solar
and the lower series to
3.0~$\times$~solar. }
\end{figure*}
%
%
%
\subsection{The `temperature problem'}
Despite the general success of the approach in Sect. 5.2,
for solar (or higher than solar) abundances one problem remains.
Only the lower border of the observed temperature sensitive 
[OIII] ratios ($\lambda4363/\lambda5007$) is reached.
In Fig.\ 9 we show
various parameters that influence this ratio.

To boost it with
a high-density ($>10^5$~cm$^{-3}$) component could be possible for low
radii (Fig.\ 9a), but would contradict the
constraint set by [OI] which would then be significantly
overpredicted.

As shown in Fig.\ 9b, matter-bounded clouds
sufficiently strengthen the [OIII] ratio only for rather thin clouds with
low surface brightness that are not very likely to dominate
(see the discussion in Sect.\ 4.2.5) and, in addition, they lead to
an overprediction of [OIII]/[OII].
 
Increasing the contribution of clouds at the lowest radius
also leads to inconsistency with [OIII]/[OII] (Fig.\ 9a).

Photoelectric
heating by dust leads to a higher electron temperature (Fig.\ 9c). However, dust
causes further effects and will  
be discussed in more detail
below (Sect. 5.4).

We are left with the reduction of the metal abundance to increase the
[OIII] ratio (Fig.\ 9d). This also improves the predictions 
of [OI]. We finally fixed the metals to $0.5\times {\rm solar}$~
except sulfur ($0.75 \times {\rm solar}$) and nitrogen (solar) because
lines from N and S ions were weakened too much by the metal reduction.
\subsection{Influence of dust}
Covering the above ranges of all parameters except abundances and adopting
the same precepts for mixing, models including dust mixed with the
emission-line gas were computed.
We adopted the Galactic-ISM typical dust composition of graphite and
silicate 
incorporated in {\em Cloudy}, and including the correspondingly depleted ISM elemental
abundances (Cowie \& Songaila 1986).

We found some favorable properties of the
inclusion of dust like an
enhancement of [NeV] (by factors $\la 3$) and HeII ($\la 2$) (this is,
however, not sufficient to obliviate the need of a black body component).
CIV is weakened (by $\la 3$) below the observed range while CIII] is
strengthened in a way to better match the observations. 
(More detailed
results are shown in Komossa 1993.)

The major deficiency of these models lies in the prediction of negligible
intensities of all Fe lines due to the strong abundance depletion of Fe. 
This includes [FeIII]$\lambda4658$~ and
[FeVII]$\lambda6087$ which are believed to be formed in the `regular'
emission-line clouds. Lines from higher ionization species might be argued
to essentially arise in a dust free extended 
component (e.g. Korista \& Ferland 1989).

We conclude that models including dust of {\em Galactic ISM} 
properties 
and correspondently modified chemical abundances 
are not able to account for the observed
Seyfert\,2 spectra. Whether other dust compositions (e.g. mainly graphite to
avoid the strong Fe depletion) could be successful is beyond the scope of
the present paper;  
too many new free parameters are introduced in this case, 
to warrant a more detailed study at the present time.
\subsection{Influence of cosmic radiation}
The processes of heating and ionization by relativistic electrons in
cosmic rays (protons in the usual energy range are much less efficient
due to their large mass)
are included in {\em Cloudy} (cf. Ferland \& Mushotzky 1982). Adding
these effects in the models by assuming the standard estimate for the
Galactic density of cosmic ray
electrons $n\tief{cr} = 2 \times 10^{-9}$~cm$^{-3}$~ leads to 
negligible changes of the line ratios ($< 1\%$).
\subsection{The `final' model series}
The dust free metal--depleted series of multi-cloud models (Sect. 5.3), 
in which the EUV continuum varies by the percentage
contribution of a 200\,000 K black body added to a steep powerlaw, 
is taken as the basis for the discussion given below. This
model sequence essentially accounts for the observed 
ranges of the emission-line intensities 
in the diagnostic diagrams (Fig.\ 10,11).
%
\begin{figure*}
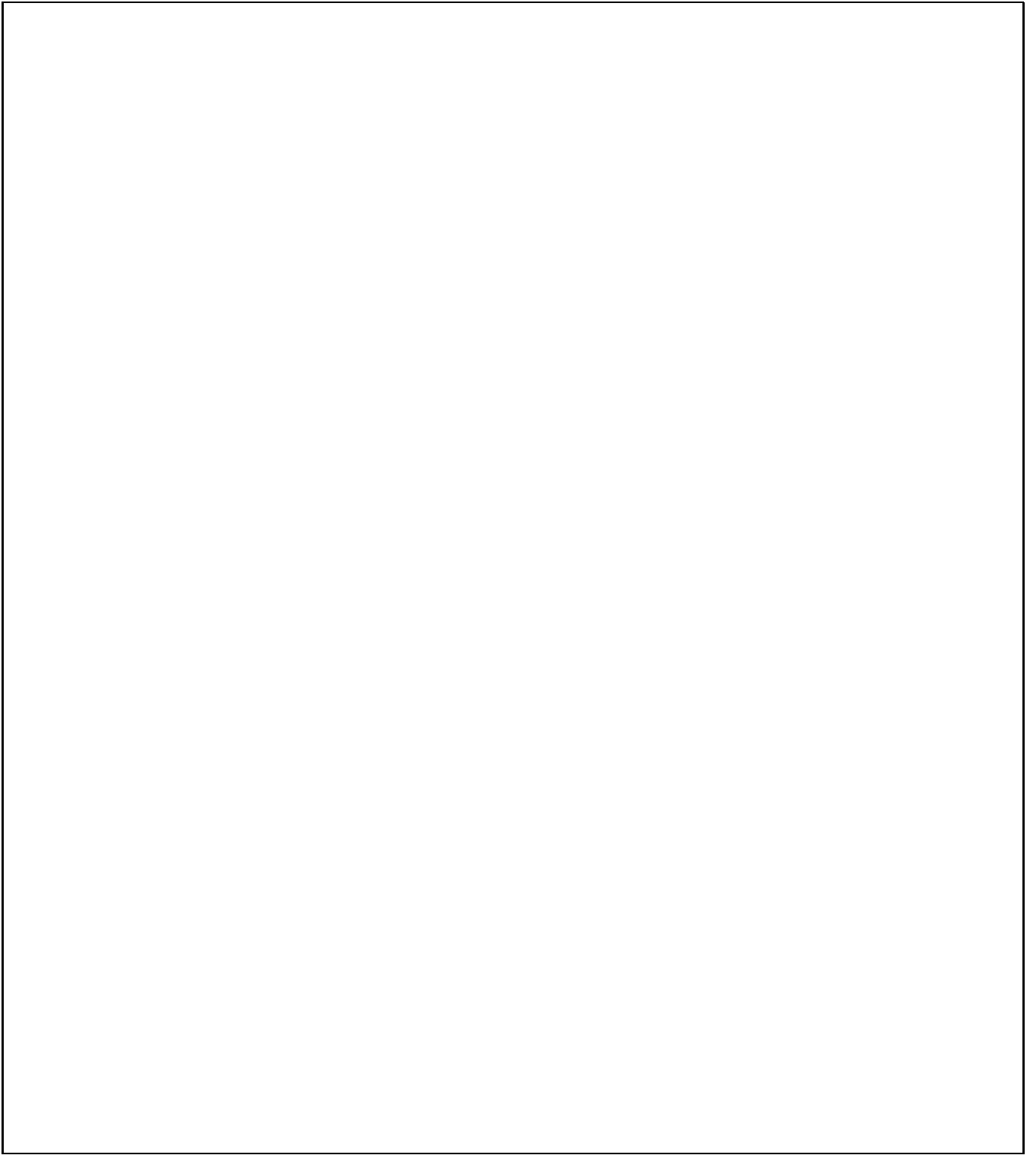

\picplace{20.3cm}
\caption[ ]{{\bf a--f}) Comparison of the observations (open symbols) with 
multi-component models (fat symbols). The metals are reduced to
0.5~$\times$~ solar. The models correspond to fixed mixtures over radius
and density (Sect. 5.1). The four asterisk symbols in each diagram
correspond to 0, 20, 50 and 100\% contributions in $Q$~ of a 
200\,000 K black body (from bottom
to top and left to right, respectively) added to an $\alpha_{\rm uv-x}=-2.0$~ EUV powerlaw
continuum. 
Fat triangles give the line ratios for powerlaw continua with
$\alpha\tief{uv-x}=-2.0,-1.5,-1.0$~ (from bottom to top, respectively) and fat dots
for black-body continua with $T\tief{bb} = 100\,000$~ and 250\,000 K
(bottom to top) for comparison. The dotted line shows the effect of the increase  
of the density from log $n_{\rm H}$ = 2..6 (direction indicated by arrow) for the fixed radius mixture and the
standard continuum ($\alpha\tief{uv-x} = -1.5$).
 }
\end{figure*}
%
\setcounter{figure}{9} 
\begin{figure*}
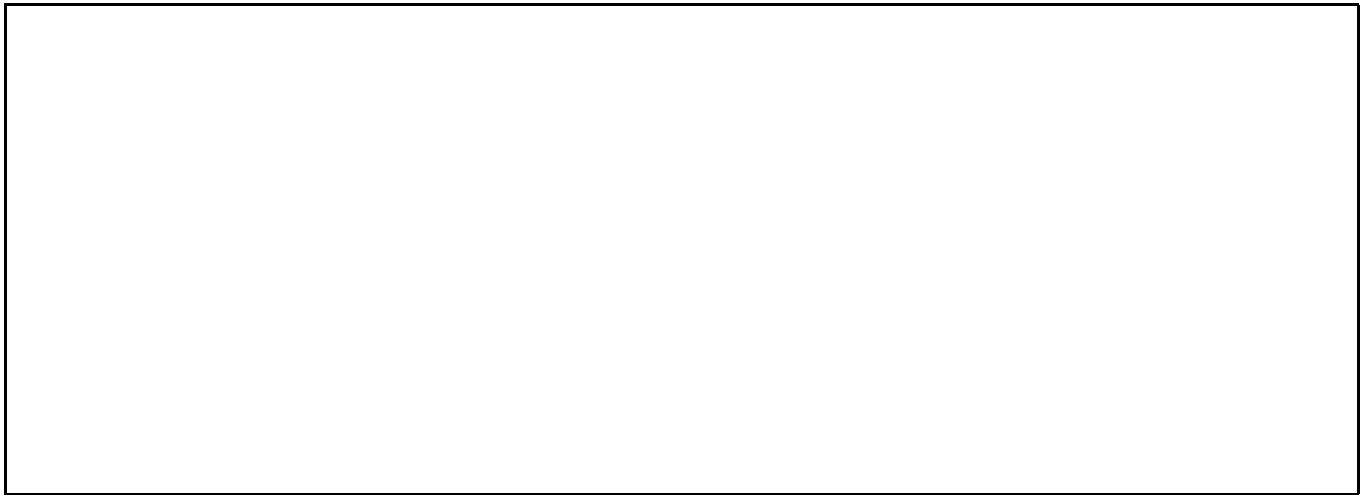

\picplace{6.5cm}
\caption[ ]{{\bf g}) Continuation of Fig. 10. 
In this diagram we also show the original (not reddening corrected)
data by small open symbols. The additional curves in this graph correspond to a range of
models including dust as marked in the figure (P$a$ = powerlaw with index $\alpha_{\rm uv-x} = -a$,
B$n$ = black body with $T = n \times 10^{4}$ K, 
PB = combination of P2 and B20 with a 50\% contribution of each component).
{\bf h}) NIR emission lines. In addition to the final model series, 
the effect of varying the chemical composition is shown (dashed lines),
with 
metal abundances of (from top to bottom) (3.0, 1.0, 0.3) $\times$ solar.  
 }
\end{figure*}
\begin{table}  
\caption{Emission line ratios $I$/$I_{\rm H\beta}$ for the final model sequence 
         including those 
         lines not shown in diagnostic diagrams, compared with the mean observed values 
         $<$obs$>$.   
         The abbreviations mark the form of the ionizing continuum: BB\,20 = black body
         with $T = 20 \times 10^{4}$ K, PL\,2 = powerlaw with index $\alpha_{\rm uv-x} = -2$,
         PB\,50 = combination of PL\,2 and BB\,20 with a 50\% contribution of each component. 
         Note that not all objects contribute to each of the listed emission lines since
         some lines are only measured for a subsample; this particularly holds for
         CIV and CIII]. Also note that [NeV] and [ArIII] are at the borders
         of the range, in which optical spectra are taken. 
         The listed observed values for all these
         lines are most probably biased
         towards high values.  
          }
     \label{tab_fin}
      \begin{tabular}{lccccc}  
    \hline
    \noalign{\smallskip}
      line & $<$obs$>$ & PL\,2 & PB\,50 & BB\,20 \\ 
    \noalign{\smallskip}
    \hline
    \hline 
    \noalign{\smallskip} 
      ~CIV 1549 & 12.0 & 1.0 & 3.8 & 7.8 \\ 
      ~CIII] 1909 & 5.5 & 0.4 & 1.3 & 2.2 \\ 
      ~MgII 2798 & 1.8 & 1.5 & 2.9 & 3.1 \\ 
      ~[NeV] 3426 & 0.9 & 0.07 & 0.4 & 0.8 \\ 
      ~[OII] 3727 & 3.3 & 1.5 & 2.5 & 3.1  \\ 
      ~[NeIII] 3869 & 1.2 & 0.6 & 1.1 & 1.4 \\ 
      ~[SII] 4074 & 0.25 & 0.2 & 0.3 & 0.4 \\ 
      ~[OIII] 4363 & 0.2 & 0.05 & 0.1 & 0.2 \\ 
      ~HeII 4686 & 0.25 & 0.06 & 0.2 & 0.35 \\ 
      ~H$\beta$ 4861 & 1.0 & 1.0 & 1.0 & 1.0 \\ 
      ~[OIII] 5007 & 9.1 & 6.2 & 10.2 & 11.5 \\ 
      ~[NI] 5200 & 0.1 & 0.1 & 0.1 & 0.1 \\ 
      ~HeI 5876 & 0.1 & 0.1 & 0.1 & 0.1 \\ 
      ~[FeVII] 6087 & 0.06 & 0.01 & 0.05 & 0.1 \\ 
      ~[OI] 6300 & 0.5 & 0.3 & 0.6 & 0.8 \\ 
      ~H$\alpha$ & 3.1 & 2.7 & 2.8 & 2.9 \\ 
      ~[NII] 6583 & 2.9 & 1.7 & 2.5 & 3.0 \\ 
      ~[SII] 6724 & 1.5 & 0.9 & 1.5 & 1.6 \\ 
      ~[ArIII] 7135 & 0.3 & 0.1 & 0.1 & 0.1 \\ 
      ~[OII] 7325 & 0.3 & 0.1 & 0.2 & 0.3 \\ 
      ~[SIII] 9069 & 0.5 & 0.3 & 0.4 & 0.4 \\ 
      ~[SIII] 9532 & 1.3 & 0.9 & 0.9 & 1.0 \\ 
    \noalign{\smallskip}
    \hline
    \noalign{\smallskip}
    \end{tabular}
  \end{table}   
\section{Discussion}
After first discussing selected diagnostic diagrams we shall deal with
the implications of the
final model sequence.
\subsection{The diagnostic diagrams}
\paragraph{The main diagrams.}
Fig.\ 10 shows that the model sequence encompasses the observed ranges in the
Veilleux-Osterbrock (1987) diagrams, i.e.  
[NII], [SII] and [OI] versus [OIII], and that it follows the main trends in
HeII and [NeIII] versus [OIII].   

Frequently, previous photoionization models encountered the difficulty of
predicting too large [OI] intensities. For two objects,
Viegas \& Prieto (1992)
solved this problem by truncating the column density of the clouds.
Whereas this might be true for a few individual Seyferts (thereby explaining
some of the observed scatter in [OI]) it seems improbable that 
in all objects the cloud-column densities adjust   
in such a way as to exactly match [OI]. 
\paragraph{[OIII] ratio versus [OIII]/[OII].}
Despite the 0.5 times reduction of the metal abundances 
the largest values of the
[OIII]$\lambda4363/5007$~ratio cannot be reached. This would be possible
with a metal depletion to 0.1 of the solar value, which, however, would
call for a supposedly too strong tuning of the N and S abundances. 
However, the
weak line [OIII]$\lambda4363$~ is rather sensitive to measurement inaccuracies
(note that five of the large [OIII] ratios plotted in Fig.\ 9 are denoted
as uncertain by the observers).  
Furthermore, doubts on the reliability of older [OIII]$\lambda$4363 
intensity measurements have recently arisen.  
Storchi-Bergmann et al. (1996) showed that 
the observed [OIII] tends to be {\em over}estimated due to improper
subtraction of the galactic stellar spectrum.  

The model sequence does not span the whole range in [OIII]/[OII]. This is a
consequence of our assumption of a fixed cloud system exposed to a
continuum sequence that only varies in shape, but with constant ionizing
photon flux.  
In practice, some spread in the average ionization parameter 
$< U >$ is expected although the spectroscopic similarities 
between Seyferts and QSOs exclude gross variations in $U$ over a wide 
luminosity range. 
Part of the scatter could be explained by a variation
of the mean density (see below; Sect. 6.4). 
\paragraph{[NI]/[NII] versus [OI]/[OII].}
There is no model sequence that can explain the weakly apparent
`anticorrelation' in this diagram 
(Fig.\ 10h) and in particular the large [NI]/[NII] at low [OI]/[OII] values
(called the `[NI] puzzle' by Stasi${\acute{\rm n}}$ska (1984)). The diagram shows that 
the reddening correction has enhanced the problem.  
Although [NI]/[NII] is boosted in models including
dust, compensating effects in [OI]/[OII] lead to avoidance of the upper
left region (Fig.\ 10h). We note, however, 
that the two objects of our sample 
at the critical upper left of the diagram appear peculiar in several diagrams (see below). 

In any case, it would be desirable to have more reliable measurements
of the weak [NI]$\lambda5200$~line. The problem is on the low-ionization
side where starburst components could be involved.
\subsection{The NIR lines}
\paragraph{[SIII]$\lambda9069+\lambda9532$~ versus [OII]$\lambda7325$.}
The near-infrared [SIII] lines are a useful aid for the classification of 
emission-line galaxies (Osterbrock et al. 1992; hereafter OTV).
They might help to distinguish between different excitation mechanisms like
photoionization or shock related processes (Diaz et al.\ 1985; Kirhakos \&
Phillips 1989). OTV noted that existing photoionization models of constant
density predict these lines about a factor of three too strong. To remove
this discrepancy, they suggested the possibility
of density-mixtures, the presence of matter-bounded clouds 
or a lower abundance of sulfur. 

However, calculations
show that the effect by 
lowering the column density would be much too small. A strong decrease of 
the sulphur abundance
would lead to a prediction of too weak {\em optical} [SII] lines.
The multi-component models of the present work nicely reproduce the average of the observations.
Whereas the observed span of the strong NIR [OII]$\lambda7325$~ feature 
is well matched
by the model sequence (Fig.\ 11b), the [SIII] observations show
a larger scatter than the predictions of the final model sequence.  
\subsection{Further lines}
\paragraph{CIII]$\lambda1909$~ and CIV$\lambda1549$.}
These UV features have been measured for only six objects in our
sample. Their large strength was difficult to be reached by
previous models. Our predictions get closer
to the observations, but still have problems in reproducing
the highest observed values. Due to observational bias the few observed
large line intensities might not be typical for AGNs. 
Upper limits that
are considerably lower have recently been reported for seven radio-loud
quasars by Wills et al.\ (1993) (upper limits of 0.56 to 2.2 for 
CIII]/H$\beta$~ and 0.64 to 2.4 for CIV/H$\beta$). Furthermore, 
it should be noted
that the calculated lines are sensitive to uncertainties in the
treatment of the Bowen resonance-fluorescence mechanism (cf. Ferland 1993).
\paragraph{Fe lines.}
[FeIII]$\lambda4658$~ and [FeVII]$\lambda6087$~ are consistently
reproduced by our models while predicted lines from higher-ionization stages
like [FeX]$\lambda6374$, [FeXI]$\lambda7892$~ and [FeXIV]$\lambda5303$~
are far below the observed values.  
However, as suggested by Korista \&
Ferland (1989) these lines might arise in a separate inter-cloud component
of tenuous gas. 
\subsection{The scatter perpendicular to the mean model sequence}
The scatter around the final model sequence often exceeds the measurement errors and
therefore requires a separate explanation. In nearly all line-ratio diagrams
a variation of the density $n\tief{H}$~ runs perpendicular to the final model sequence.
Therefore,
a variation of the density distribution 
could account for the scatter (Fig.\ 10). 

This does, however, not
work in the NIR [SIII]-[OII] diagram  
in which the model sequence is rather
tight.               
Metal abundance variations may induce some of the scatter here
(that usually lead to changes in the same direction as the continuum shape,
although less pronounced,  
in most other diagrams). 

We note that the original data (uncorrected for reddening) usually show
the same trends in the line-ratio diagrams, i.e. our main results are rather insensitive
to the exact value of reddening.   

A few objects are conspicuous in the diagrams because they have a few exceptional
line ratios that do not fit to the rest of the sample. E.g., Mrk 273 (No. 4 in
Fig. 1) is exceptionally weak in [OIII] and exceptionally strong in [OII]
and peculiar in further line ratios. NGC 1068 (No. 32) is
exceptionally strong in [NII]. 

The presence of a contaminating starburst or
HII component cannot be excluded in all cases, especially for the low
ionization objects. So, also [NI] might be strengthened in a dusty
starburst.

With these possibilities in mind it is not warranted to further tune
our final model sequence which is remarkably successful, especially for
normal to high ionization Seyfert\,2s.
\subsection{Implications of the multi-component model}
The `final' Seyfert\,2 model consists of a radially distributed
ensemble of clouds with densities $n\tief{H} = 10^2$~ to $10^5$~cm$^{-3}$
and subsolar (except nitrogen) metal abundances exposed to the ionizing
radiation of a continuum source in the center of the galaxy. The
EUV continuum is characterized by a steep powerlaw plus a bump component,
parameterized as 
hot black body the strength of which varies from
object to object.  
Two questions arise: First, is this model in accordance with other
observational evidence? Second, is there any model that would
explain how the derived structure of the NLR could have formed and
how it can be pertained? Having these questions in mind we
in turn discuss the topics cloud origin, cloud confinement, and 
abundances.

\paragraph{Origin of the NLR gas, density range and cloud confinement.}
It is unclear how the NLR emission-line clouds are formed and how they are 
confined. 
Since the 
line widths suggest a relation to the dynamics of bulge stars 
(e.g. Wilson \& Heckman 1985, Terlevich et al. 1990, Nelson \& Whittle 1996)
mass-loss of bulge stars might have delivered the material 
for the NLR clouds. 

By noting that Seyfert nuclei host galaxies are early-type spirals, i.e. with large
bulges, early speculations (e.g. Schulz 1986) suggested as a second 
possibility accretion of a dwarf galaxy. 

In a third scenario proposed 
by Cameron et al.\ (1993) the NLR clouds arise from gas ablated from
molecular-cloud complexes. 

Concerning the density structure of the NLR, there might even be a density stratification
within a single ionization-bounded entity identified as a `cloud' but the lack of a
detailed theory for such a structure hampers the application of appropriate
photoionization models.
Our approach of employing different clouds each of constant density
represents an approximation of the possibly more complicated real situation.

Photoionized clouds at the same radius but with different densities will attain
quite different pressures so that a confinement problem may occur. For
cloud thicknesses of the order of $10^{19}$~ to $10^{21}$~ cm sound
crossing times are in the range $\sim 10^{5-7}$~ years, which is on the
orbital timescale in which destabilizing effects
are likely to operate (cf. Mathews \& Veilleux 1989). If clouds are destructed
on short timescales one needs comparatively fast replenishment.
The three mechanisms mentioned above
could lead to a steady supply of NLR gas for more than
$10^8$ years so that requirements for cloud-confinement mechanism
would be much relaxed. 

\paragraph{Radial extent.}
The radial extend of the NLR gas refers to a model with fixed $Q$. Since
the observed objects scatter in $Q$, our approach implies a scaling of
NLR radius with object luminosity, similar to what is directly observed
for the BLR (e.g. Peterson 1993).

\paragraph{Abundances.}
It is unclear which abundances are to be expected
and a priori, there is no reason to expect them
to be exactly solar. The question on the abundances is closely linked to 
the origin of the NLR gas.  
Our final model requires subsolar metal abundances of about 
0.5 times the solar value.  

Abundance determinations made with `nebular methods' in
HII regions within the spiral arms of a few Seyferts 
showed no strong differences to normal galaxies (Evans \& Dopita 1987).
On the other hand, Balcells \& Peletier (1994) found subsolar metal abundances in spiral bulges.
This would fit to the possible stellar origin of the NLR gas in Seyferts. 
Depleted gas-phase oxygen abundances were also found
by Sternberg et al. (1994) in molecular gas in the central region of NGC 1068
and material evaporating from the torus may be an alternative reservoir for NLR clouds.  
In the same object, Marshall et al. (1993) found evidence for
underabundant oxygen of 1/5 $\times$ solar in the warm electron scattering medium
using X-ray data.   
Recently,
in a large sample of spiral galaxies Zaritsky et al.\ (1994)
derived a variety of O/H abundance ratios between 0.26 and 2.4 times the solar
value. 

\paragraph{Velocity field.}
In our final NLR model, 
high-ionization lines preferentially originate in clouds near the nucleus, whereas 
the major contribution to more lowly ionized species comes from further outwards. 
Consequently, a trend for a correlation of line-width with ionization potential 
is expected in case of decelerated outflow, accelerated inflow or rotation
of the NLR material, whereas all lines are expected to have comparable line-widths 
in case of a dominating turbulent velocity component. 

Whittle (1985) studied the line width variation versus ionization
potential and critical density in a few Seyfert\,2s and found only weak trends with
no clear preference for either abscissa.
This calls for a strong `turbulent' (locally random velocity field) component, 
comparable to that found via a line-profile analysis e.g. in NGC 4151 
(Schulz 1989, 1990).  

\paragraph{EUV bump.}
One of the most remarkable features of the final model sequence
is the varying dominance of the hot           
black body in the ionizing EUV.
Although slightly narrower, this component might mimic the thermal component 
of an accretion disk.
Another possibility would be that the black body crudely
simulates a free-free component for $T < 10^6$ K.

In the IUE-UV of Seyfert\,2s there is no indication
for a rise towards shorter wavelength, so one might
infer that the dominant temperature of a significant
bump component must indeed be high. 
However, the absolute UV brightness of Seyfert\,2s amounts only 10\% 
of that of a typical Seyfert\,1,  
so that rather than seeing
the continuum source directly we are likely to see a scattered component in agreement
with the unified model (e.g. Mulchaey et al.\ 1994). Adopting the `unified view' that Seyfert\,2s
are more inclined Seyfert\,1s it is justified to make a comparison with Seyfert\,1 continua.
Recent quasi-simultaneous UV -- soft X-ray studies (Walter \& Fink 1993; Walter et al.\ 1994)
of Seyfert\,1 galaxies indeed suggest the presence of an EUV component of varying strength but
constant shape. 

Independent evidence for the existence of EUV bumps comes from the so-called photon-deficit problem,
one interpretation of which is the existence of an additional EUV bump component (e.g. Penston et al. 1990,
Kinney et al. 1991). 
\section{Summary and conclusions}
Emission-line spectra from 37 Seyfert\,2 galaxies are utilized to find constraints
for photoionization models that are required to fit all available lines. 
We discuss the influence of several parameters as are the radial extent of
the NLR, the gas density, the cloud column density, the metal abundance, dust and
the shape of the ionizing continuum. The results are used to  
build a multi-component NLR model, 
thereby solving or diminishing known problems in previous photoionization approaches, 
and to filtrate the parameter that governs the trends and correlations in the
emission-line diagrams.   

To explain both high and low ionization
lines, emission-line gas has to be
distributed over a range of galactocentric radii 
(of $\sim $ $10^{20} \sqrt{Q_{\rm tot}/10^{54}\rm s^{-1}}$ cm to $10^{21.5}
\sqrt{Q_{\rm tot}/10^{54}\rm s^{-1}}$~ cm). 
Furthermore, no single
hydrogen density can account for all lines. The range covers
$n\tief{H} = 10^2$~ to $10^5$~ cm$^{-3}$. 
Metal abundances must be subsolar ($\sim 0.5 \times$ solar) for 
at least all 
objects with a measured [OIII]$\lambda4363/\lambda5007$~ line ratio. 
This is found to be the only consistent solution to the `temperature problem'
for our sample, as compared to the possibility of 
the significant contribution of a high-density component ([OI] strongly overpredicted),
the presence of dust internal to the NLR clouds (Fe lines underpredicted), or
the contribution of matter-bounded clouds ([OIII]/[OII] overpredicted).  

A {\em systematic} variation of the cloud column densities
was found to be unsuccessful in explaining the trends and correlations 
in the diagnostic diagrams. Neither can it account for the clear
correlation of e.g. [NeIII] with HeII, nor for the fact that [NI] is
not clearly correlated with e.g. [OI] and [SII]. 
We cannot exclude a random contribution of matter-bounded clouds in
individual objects, however. 

We investigated powerlaw ($\alpha\tief{uv-x} = -2.5$~ to $-1$) 
and hot black-body components
($T\tief{bb} = 100\,000 \mbox { to }250\,000$~ K) in the EUV continuum. 
The highest values of HeII, [NeV] and CIII] can only be explained
by including significant EUV bumps parameterized as black bodies. Irrespective of other parameters, 
a single fixed continuum shape cannot explain the total range of Seyfert\,2 line
ratios.

As a consequence of these results, we devised the following
multi-component approach.
The NLR is composed of an ensemble of ionization-bounded, metal-depleted
clouds distributed over a range of radii, with a radius-independent mixture 
of densities. 
Along the model series in the emission-line diagrams (i.e. from one object
to another), only the shape of the ionizing EUV continuum is varied. 
It consists of a steep powerlaw with
$\alpha\tief{uv-x} \approx -2$~ plus a hot black body with
a temperature $T\tief{bb} \approx 200\,000$~K. The contribution of $Q\tief{bb}$~
to $Q\tief{tot} = Q\tief{pl} + Q\tief{bb}$~ spans 0 to 100\%.

This model scheme matches the mean Seyfert\,2 line ratios and reproduces the
basic correlations. Deviations of individual objects from the mean can be
introduced by, e.g., density variations. 
\acknowledgements{We are indebted to Gary Ferland for providing
{\em Cloudy}. H.S. acknowledges support by the German Space Agency
DARA under 50 9102.}

 \end{document}